\documentclass[10pt,amsmath,amssymb,nofootinbib,twoside,twocolumn,superscriptaddress,floats,floatfix,aps,prd,preprintnumbers]{revtex4-2}
\usepackage[british]{babel}
\usepackage{xcolor}
\usepackage{booktabs}
\usepackage{latexsym}
\usepackage{amssymb}
\usepackage{subfigure}
\usepackage{amsmath}
\usepackage{graphicx}
\usepackage{tabularx}
\usepackage{cancel}
\usepackage{hyperref}
\hypersetup{
    colorlinks=true,
    linkcolor=black,
    citecolor=blue,
    filecolor=black,
    urlcolor=blue,
    pdfauthor={M. Cadoni, M. Pitzalis, and A. P. Sanna},
    pdftitle={Apparent horizons in cosmologically-embedded black holes}
}
\usepackage[capitalise]{cleveref}
\usepackage[utf8]{inputenc}
\usepackage[normalem]{ulem}
\def\beq#1\eeq{\begin{align}#1\end{align}}

\newcommand{\dd}{\text{d}}

\newcommand{\rAH}{r_\text{AH}}
\newcommand{\rH}{r_\text{EH}}

\begin{document}

\newcommand{\UniCa}{\affiliation{Dipartimento di Fisica, Universit\`a di Cagliari, Cittadella Universitaria, 09042 Monserrato, Italy}}
\newcommand{\INFNCa}{\affiliation{INFN, Sezione di Cagliari, Cittadella Universitaria, 09042 Monserrato, Italy}}

\author{M.~Cadoni}
\email{mariano.cadoni@ca.infn.it}
\UniCa\INFNCa

\author{M.~Pitzalis}
\email{mirko.pitzalis@ca.infn.it}
\UniCa\INFNCa

\author{A.~P.~Sanna}
\email{asanna@dsf.unica.it}
\UniCa\INFNCa

\title{Apparent horizons in  cosmologically-embedded  black holes}

\begin{abstract}
     We present exact solutions for the cosmological embedding of a broad class of non-singular black holes, demonstrating that these objects exhibit an apparent horizon. The evolution of the latter is analyzed as a function of the cosmological redshift $z$. We show that its size exceeds that of the event horizon of an isolated black hole and increases monotonically with increasing $z$. Explicit formulas and numerical results are provided for the specific cases of the Hayward and Fan $\&$ Wang non-singular black-hole models. Furthermore, we explore the distinct dynamical roles of the event and apparent horizons, highlighting connection between the latter and the recently identified cosmological mass shift in non-singular black holes. 
\end{abstract}

\maketitle

\section{Introduction}
The possibility of a dynamical coupling between local astrophysical objects, such as black holes, and the large-scale cosmological dynamics within the framework of General Relativity (GR) is an old issue. It was first discussed by McVittie~\cite{McVittie:1933zz},  subsequently developed by Einstein and Straus~\cite{Einstein:1945id,Einstein:1946zz}  and  recently has gained considerable attention and interest~\cite{Croker:2024jfg,Faraoni:2023hin,Calza:2024qxn,Calza:2024xdh,Calza:2024fzo,dicke1964evolution,Vaidya:1968zza,DEath:1975jps,
Gautreau:1984pny,Cooperstock:1998ny,Nayak:2000mr,Baker:2000yh,Bolen:2000dz,Dominguez:2001it,Ellis:2001cq,Gao:2004cr,Sheehan:2004wa,
Nesseris:2004uj,Sultana:2005tp,Li:2006zh,Adkins:2006kw,McClure:2006kg,Sereno:2007tt,Faraoni:2007es,Balaguera-Antolinez:2007csw,Mashhoon:2007qm,Carrera:2008pi, Gao:2011tq,Faraoni:2014nba,Kopeikin:2014qna,Faraoni:2015saa,Mello:2016irl,Faraoni:2018xwo,Guariento:2019ock,Spengler:2021vxy,
Agatsuma:2022ewd,Croker:2019mup,Croker:2020,Croker:2020plg,Farrah:2023opk,Wang:2023aqe,Gaur:2023hmk,Parnovsky:2023wkc,Avelino:2023rac,Dahal:2023hzo,Gao:2023keg,Cadoni:2023lum,Cadoni:2023lqe,Cadoni:2024jxy}. The primary outcome of these recent theoretical investigations is the prediction of a power-law relation between the masses of local compact objects and the cosmological scale factor $a$, expressed as~\cite{Croker:2019mup}:
\begin{equation}
\label{msf}
    M(a) \propto a^k, \qquad -3\le k\le 3\, .
\end{equation}
Additionally, the exact form of $M(a)$ has been derived within a solid GR framework~\cite{Cadoni:2023lum,Cadoni:2023lqe,Cadoni:2024jxy}. For black holes, this framework predicts the scaling \eqref{msf}, with $k=0$ for singular  black holes or $k=1$ for non-singular ones~\cite{Cadoni:2023lum,Cadoni:2023lqe}.   

While the observational determination of $k$ is a complex and controversial issue~\cite{Farrah:2023opk,Cadoni:2023lum,Rodriguez:2023gaa,Andrae:2023wge,Lei:2023mke,Amendola:2023ays,Lacy:2023kbb,Calza:2024qxn}, the very concept of black-hole cosmological coupling has faced skepticism. The main objection is the significant separation of scales between local and cosmological structures, which would seemingly make such interactions physically implausible~\cite{Wang:2023aqe,Gaur:2023hmk}. However, scale separation in a cosmological background is a nuanced concept, well-defined only for spacetimes like Schwarzschild-de Sitter, which allows for a static parametrization~\cite{Cadoni:2023lqe}. 

A recent study~\cite{Faraoni:2024ghi}, approaching the issue from a different perspective, argues that the cosmological coupling of black holes is unavoidable. Without relying on Einstein's equations or specific matter content, the study shows that the presence of an event horizon (EH) is physically untenable for an exactly static, spherically-symmetric black-hole horizon embedded in a cosmological background, as it would result in a naked singularity at the would-be horizon. Instead, a time-dependent apparent horizon (AH) seems more appropriate for characterizing the cosmological embedding of local, highly-compact objects. While the findings of Ref.~\cite{Faraoni:2024ghi} strongly support cosmological coupling for black holes, they also raise several new questions. 

First, can the existence of cosmological AHs serve as a proxy for cosmological mass coupling, as described by \cref{msf}? The answer is not a straightforward “yes” due to counterexamples like the McVittie spacetime, which represents the cosmological embedding of a Schwarzschild black hole. While its AH evolves with time~\cite{Faraoni:2012gz}, this solution does not exhibit cosmological mass coupling~\cite{Cadoni:2023lqe}.

Second, is there a relationship between the AH and the EH of an isolated black hole, as observed locally at a fixed cosmological time? This question is quite intricate and cannot be resolved merely by the analogy with event and apparent horizons in collapsing bodies. In the latter case, an EH is globally defined as it requires knowledge of the infinite future of the collapsing body, and only exists in static spacetimes. Conversely, an AH is locally defined and does not depend on the entire spacetime history. Moreover, in our cosmological context, the notions of event and apparent horizons apply to two distinct regimes. The EH characterizes the static, eternal black hole accessible to local observers, while the AH is a hallmark of the cosmological embedding.

Answering these questions is challenging, particularly because we currently lack explicit examples of cosmologically-embedded black-hole solutions (CEBHSs), aside from the McVittie or Schwarzschild-de Sitter solutions.

The main goals of this paper are to derive explicit CEBHSs, determine the time evolution of their AH and address the previously-raised questions. We start by constructing explicit exact solutions for the cosmological embedding of a broad class of non-singular black holes proposed in Ref.~\cite{Cadoni:2022chn} (\cref{sect:2}). Using these solutions, we derive the equation governing the time evolution of the AH (\cref{sect:3}). We also provide detailed analysis of two particular cases of interest: the cosmological coupling of Hayward~\cite{Hayward:2005gi} and Fan $\&$ Wang~\cite{Fan:2016hvf} regular black holes (\cref{sect:4}). These results will allow us to provide answers to the previously raised questions (\cref{sect:5}).

\section{Cosmologically-embedded non-singular black-holes: Exact solutions}
\label{sect:2}

A cosmologically-embedded black hole is a solution to Einstein's equations that depends on both the radial coordinate $r$ and conformal time $t$. This solution describes an isolated black hole at fixed $t$, while asymptotically approaching Friedmann-Lema\^itre-Robertson-Walker (FLRW) cosmology at large $r$. 
Focusing on static, spherically-symmetric black holes, the spacetime metric can be generally parametrized as follows~\cite{Cadoni:2023lum,Cadoni:2023lqe} \footnote{Unless otherwise stated, we adopt units in which $c = \hbar = k_\text{B} = 1$.}:
\begin{equation}
\dd s^2 = a^2(t)\left[-e^{\alpha(t, r)} \dd t^2 + e^{\beta(t, r)} \dd r^2 + r^2 \dd\Omega^2\right]\, .
\label{generalmetric}
\end{equation}
Here, $\alpha(t, r)$ and $\beta(t, r)$ are metric functions, depending on both the conformal time $t$ and the radial coordinate $r$, while $a(t)$ is the cosmological scale factor.

Constructing CEBHSs is quite challenging. McVittie~\cite{McVittie:1933zz} used a "cosmological uplifting method" to embed a Schwarzschild black hole in a cosmological background. With this method, static solutions are promoted to time-dependent ones, while Einstein's equations determine the energy density and pressure required to sustain the solution. In its original form, the McVittie method only applies to the Schwarzschild black hole, due to its reliance on isotropic sources.

Recent studies~\cite{Cadoni:2023lum, Cadoni:2023lqe,Cadoni:2024jxy} have emphasized the crucial role of anisotropies in constructing CEBHSs.
The reason is that with isotropic pressure, Einstein's equations and conservation laws often form an over-constrained system. Anisotropic fluids, however, introduce two independent pressure components, providing additional degrees of freedom to construct CEBHSs~\cite{Cadoni:2023lum, Cadoni:2023lqe, Cadoni:2024jxy}. 
Furthermore, anisotropic sources represent the simplest way to circumvent Penrose's singularity theorem, enabling the construction of regular black-hole solutions~\cite{Cadoni:2022chn}. For these reasons, we consider sources described by an   anisotropic fluid, 
whose energy-momentum tensor is given by
\begin{equation}
\label{TensoreEIIntro}
    T_{\mu\nu} = \left(\epsilon + p_{\perp} \right)u_{\mu}u_{\nu} + p_{\perp} \ g_{\mu\nu} - \left(p_{\perp}-p_{\parallel} \right)w_{\mu}w_{\nu}\, ,
\end{equation}
where $\epsilon$, $p_\parallel(r)$ and $p_\perp$ are the energy density and pressure components, respectively, while $u_\mu$ and $w_\mu$ are a timelike and spacelike 4-vectors, respectively, satisfying the normalization conditions $g^{\mu\nu} u_{\mu}u_{\nu} = -1$, $g^{\mu\nu} w_{\mu}w_{\nu} = 1$ and $u^{\mu}w_{\mu} = 0$. The isotropic case is trivially recovered in the limit $p_\parallel = p_\perp = p$.

Refs.~\cite{Cadoni:2023lum,Cadoni:2023lqe} demonstrated that this setup, with $\dot \alpha = 0$, allows to describe the embedding of black holes/compact objects in a cosmological background. Einstein's equations determine the $g_{rr}$ component of the metric in terms of the static profile $\beta_0(r)$ at a fixed reference time $t_0$ (typically taken as the present time)~\cite{Cadoni:2023lum, Cadoni:2023lqe, Cadoni:2024jxy}, as follows
\begin{equation}
    e^{-\beta(r, t)} = e^{-\beta_0(r)} a(t)^{k(r)}\, , \qquad k(r) \equiv r \alpha'(r)\, .
    \label{betafunction}
\end{equation}
In the limit $r \to \infty$, the large-scale, cosmological dynamics, encoded in the scale factor $a$, completely decouples and it is described by the standard Friedmann equations (in a spatially-flat universe)
\begin{subequations}
\begin{align}
&3\frac{\dot a^2}{a^2} = 8\pi G a^2 \epsilon_1(t) \, ; \label{Fried1}\\
&\frac{\dot a^2}{a^2}-2\frac{\ddot a}{a} = 8\pi G a^2 p_1(t)\, ; \label{Fried2}\\
&\dot \epsilon_1 + 3 \frac{\dot a}{a}\left(\epsilon_1 + p_1 \right) = 0\, , \label{conscosmo}
\end{align}
\label{systemFried}
\end{subequations}
where $\epsilon_{1}(t)$ and $p_{1}(t)$ are the cosmological time-dependent energy density and pressure, respectively. 

Instead, the local, static, spherically-symmetric solution is obtained from the general equations evaluated at fixed time $t_0$
\begin{subequations}
\label{SystemEqFirstRegime}
\begin{align}
&\frac{1-e^{-\beta_0}+r\beta_0'e^{-\beta_0}}{r^2} = 8\pi G \, \epsilon(r)\, ; \label{alpha00static}\\
&\frac{e^{-\beta_0}+re^{-\beta_0} 	\alpha'-1}{r^2}=8\pi G  \, p_{\parallel}(r)\, ; \label{alpharrstatic} \\
&p'_{\parallel} +\frac{\alpha'}{2}\left(\rho+p_{\parallel}\right)+ \frac{2}{r}\left(p_{\parallel}-p_\perp\right)=0 \label{press1}\, .
\end{align}
\end{subequations}
In the above, $\epsilon(r)$, $p_{\parallel}(r)$ and $p_\perp(r)$ are the time-independent density and pressure profiles of the local object at the present time. The local solution is described by two independent functions $\alpha(r)$, $\beta_0(r)$ \footnote{Alternatively, the Misner-Sharp mass $M_\text{MS}$ can be used instead of $\beta_0(r)$, with the relation $e^{-\beta_0(r)} = 1-{2GM_\text{MS}(r)}/{r}$.}. Once these functions are determined, the full cosmologically-embedded solution is obtained by using \cref{betafunction} to get $\beta(t,r)$. The time-dependent pressures $p_{\parallel}(t,r)$, $p_\perp (t,r)$ and density $\epsilon(t,r)$ are then provided by using Einstein's equations~\cite{Cadoni:2023lum,Cadoni:2023lqe,Cadoni:2024jxy}.

In particular, for a non-singular black hole, integrating in the comoving volume the time-time component of Einstein's equations, giving $\epsilon(t,r)$, yields the Misner-Sharp mass and the mass-shift formula \eqref{msf}, with $k=1$~\cite{Cadoni:2023lum}
\begin{equation}
M_\text{MS}(a)=  M_\text{MS}(a_i) \frac{a}{a_i}\, ,
\label{mshift}
\end{equation}
where $a_i$ is the scale factor at the time of the object's formation.
The explicit forms of $\beta_0(r)$ and $\alpha(r)$ depend on the specific black-hole being embedded. Here, we focus on the general class of regular black-hole models analyzed in~\cite{Cadoni:2022chn}. These models describe regular black holes with a de Sitter core, that asymptotically approach the Schwarzschild behavior. This class includes, for instance, quantum-deformed Schwarzschild black holes (see Ref.~\cite{Cadoni:2022chn} and references therein), effective models like, e.g., the Hayward and Bardeen black holes~\cite{bardeen1968proceedings,Hayward:2005gi} or black holes sourced by non-linear electrodynamics~\cite{Ayon-Beato:1998hmi,Ayon-Beato:1999qin,Bronnikov:2000vy}. These models are parametrized by a single metric function $f(r)$, related to the functions appearing in \cref{generalmetric,betafunction} by:
\begin{equation}
    f(r) = e^{\alpha(r)} = e^{-\beta_{0}(r)}\, .
    \label{ffunction}
\end{equation}
The function $f$ depends on two parameters, the two hairs characterizing the black hole: its mass $M$ and an additional (possibly quantum) hair $\ell$. Moreover, it must  satisfy several conditions: $(1)$ a de Sitter behavior near $r=0$, which also ensures the removal of the Schwarzschild curvature singularity; $(2)$ two horizons that degenerate at extremality. This implies the following near-horizon behavior of the metric function near the outer EH, located at $r=\rH$:
\begin{equation}
    \label{nhb}
    \alpha\sim \ln(r-\rH), \qquad %\alpha' \sim r\alpha'=k(r)\sim (r-r_H)^{-1}.
    r \alpha' = k(r) \sim r (r-\rH)^{-1}
\end{equation}
$(3)$ an asymptotic Schwarzschild behavior as $r \to \infty$
\begin{equation}
    \label{alimit}
    \alpha\to 0, \qquad r \alpha'\to 0 \, .
\end{equation}

Once a suitable profile for $f(r)$, satisfying the aforementioned constraints, has been chosen, the corresponding CEBHSs is fully described by \cref{generalmetric} with $\beta(r,t)$ determined by \cref{betafunction}. 
In the following section, we will derive the evolution of the AH for the generic solution. 
In \cref{sect:4}, we will focus on two particular and notable cases: the cosmological embedding of the Fan $\&$ Wang~\cite{Fan:2016hvf} and Hayward~\cite{Hayward:2005gi} regular black holes. 

\section{Apparent horizon}
\label{sect:3}
For spherically symmetric solutions, the location of the AH is given by the solution (if it exists) of the equation
\begin{equation}
    \nabla^a R \nabla_a R = g^{ab} \partial_a R \partial_b R = 0\, ,
\end{equation}
where $R = R(t, r)$ is the area of the $2$-sphere. Physically, this means we are imposing the spherical surface of radius $R$ to be null. In the particular case of \cref{generalmetric}, $R(t, r) = a(t) r$. This is analogous to computing the zeroes of $g^{RR} = 1-\frac{2G M_\text{MS}}{R}$. For the metric \eqref{generalmetric}, the latter reads as~\cite{Cadoni:2023lqe}
\begin{equation}\label{MSmassSchwarzschildcoordinates}
M_\text{MS} = \frac{R}{2G}\left(1 + \dot R^2 e^{-\alpha}-R'^2 e^{-\beta} \right)\, .
\end{equation}
Setting $1-\frac{2G M_\text{MS}}{R} = 0$, then \footnote{One can easily verify that, as anticipated, the following result is equivalent to $g^{ab}\partial_a R \partial_b R = g^{tt} \dot R^2 + g^{rr}R'^2 = 0$.}, and using \cref{betafunction}, the equation for the location of the AH $\rAH$ for the generic cosmologically-coupled solution reads as 
\begin{equation}
\begin{split}
    %&\frac{R}{2G} = M_\text{MS} = \frac{R}{2G}\left(1 + \dot R^2 e^{-\alpha}-R'^2 e^{-\beta} \right),\\
    %&\dot R^2 e^{-\alpha(\rAH)} = R'^2 e^{-\beta_0(\rAH)}\,  a^{\rAH \alpha'(\rAH)},\\
    %&\frac{\dot a^2}{a^2} \, \, \rAH^2 \, e^{-\alpha(\rAH) + \beta_0(\rAH)} = a^{\rAH \alpha'(\rAH)},\\
    &\frac{\dot a^2}{a^2} \, \rAH^2 \, e^{-2\alpha(\rAH)} = a^{\rAH \alpha'(\rAH)}\, ,
\end{split}
\label{apparenthorizoneq}
\end{equation}
where we used  \cref{ffunction}, $\dot R = \dot a \, r$ and $R'= a$. Using the Friedmann equation \eqref{Fried1}, the left-hand side can be recast as 
\begin{equation}
    \frac{8\pi G}{3} \epsilon_1(t) \, \rAH^2 e^{-2\alpha} = a^{\rAH \alpha' -2}\, .
    \label{apparenthorizonintermediatestep}
\end{equation}
In the following, we assume the cosmological background to be sourced by a perfect fluid with barotropic equation of state $p_1(t) = w \epsilon_1(t)$, such that \cref{conscosmo} yields 
\begin{equation}
    \epsilon_1(t) = \epsilon_0 \, a^{-3(1+w)}\, .
\label{cosmologicaldensity}
\end{equation}
Substituting \cref{cosmologicaldensity} into \cref{apparenthorizonintermediatestep} gives
\begin{equation}
\begin{split}
    %&\frac{8\pi G}{3} \epsilon_0 \, \rAH^2 \, e^{-2\alpha} = a^{\rAH \alpha'+1+3w}\\
    H_0^2 \Omega \, \rAH^2 \, e^{-2\alpha} =  \left(\frac{1}{1+z} \right)^{\rAH \alpha'+1+3w}\, ,
\end{split}
\label{ApparentHorizonFinalEq}
\end{equation}
where we used the definition of the cosmological parameter $\Omega \equiv 8\pi G \epsilon_0/3H_0^2$ and converted the scale factor to the redshift through $a = (1+z)^{-1}$.

To avoid any confusion, we stress again that in our cosmological framework, the two notions of event and apparent horizons not only have an entirely different meaning, but also correspond to two distinct physical situations. 
The EH can be defined solely for a static, eternal, black hole. Consistently with the results of Ref.~\cite{Faraoni:2024ghi}, it does not exist for cosmologically-embedded solutions, for which only the AH is relevant. Physically, this means that the black hole's EH can be detected only by a local observer at a fixed redshift. This is the same local observer who tests the mass-shift formula \eqref{mshift} by comparing local measurements of the Misner-Sharp mass across different redshifts.

\subsection{Evolution of the apparent horizon}{\label{sect:31}}
\Cref{ApparentHorizonFinalEq} encodes in implicit form the form of the function $\rAH(a)$,  which describes   evolution of the AH of our CEBHSs with the scale factor during cosmological expansion. We will first determine the qualitative form of the function $a(\rAH)$ and from that infer the form of the inverse function $\rAH(a)$.  

The qualitative behavior of the function $a(\rAH)$ can be deduced  from the general features of the metric function $f$, as discussed in \cref{sect:2}. We will focus on CEBHSs which feature an outer EH when evaluated at fixed time. 

Outside the EH $(1)$ $\alpha'$ is always positive, diverges at the horizon and  goes to zero asymptotically; $(2)$ $\alpha$ is negative near the horizon (where it tends to $-\infty$), then increases and approaches to zero asymptotically. We also assume that our black hole, with outer EH at $\rH$, forms at some redshift $z = z_i$, and remains coupled to the cosmological dynamics thereafter. We also first consider the case $w>-1/3$. 

Let us now rewrite \cref{ApparentHorizonFinalEq} in  the form 
\begin{equation}
\begin{split}
\ln a &= \frac{\ln\left[ H_0^2  \rH^2 \Omega\right] + 2\ln \left(\frac{\rAH}{\rH}\right) -2\alpha}{\omega}\, ,  \\
\omega &\equiv {\rAH \alpha'+1+3w}\, .
\end{split}
\label{ApparentHorizonFinalEq1}
\end{equation}
Outside the EH, the denominator in \cref{ApparentHorizonFinalEq1} is always positive since $\alpha'\ge 0$. In the numerator, instead, the third term is always positive, while the first one is always negative, with $-2\alpha \to \infty$ at the EH. As a result, in the near-EH region, the right hand side of \cref{ApparentHorizonFinalEq1} is positive, meaning that \cref{ApparentHorizonFinalEq1} has no solutions (we are considering $a<1$).
This implies that the minimum value of the AH, $\rAH^m$, is always larger than the EH. This minimum is reached when $a = 1$, i.e., when the numerator of \cref{ApparentHorizonFinalEq1} vanishes. $\rAH^m$ is the smallest root of the equation 
\begin{equation}
\frac{f}{r}= H_0\sqrt{\Omega} \, , 
\label{rAHm}
\end{equation}
and depends on the black-hole hairs $M$, $\ell$  entering the metric function $e^\alpha$, as well as the cosmological parameters $H_0^2$, $\Omega$. It does not depend on $w$.

As $\rAH$ increases above $\rAH^m$, the positive contribution coming from the $-2\alpha$ term diminishes, driving also the scale factor $a$ towards smaller values. At large values of $\rAH$, both $\alpha$ and $\alpha'$ approach zero, but the positive $\ln r/\rH$ term starts to dominate. Eventually, both terms cancel again the negative term $\ln[H_0  r_H^2 \Omega]$, determining a second, larger, root $\rAH^s$ of \cref{rAHm}, corresponding again to $a=1$.

\begin{figure}[!t]
    \centering
    \includegraphics[width=\linewidth]{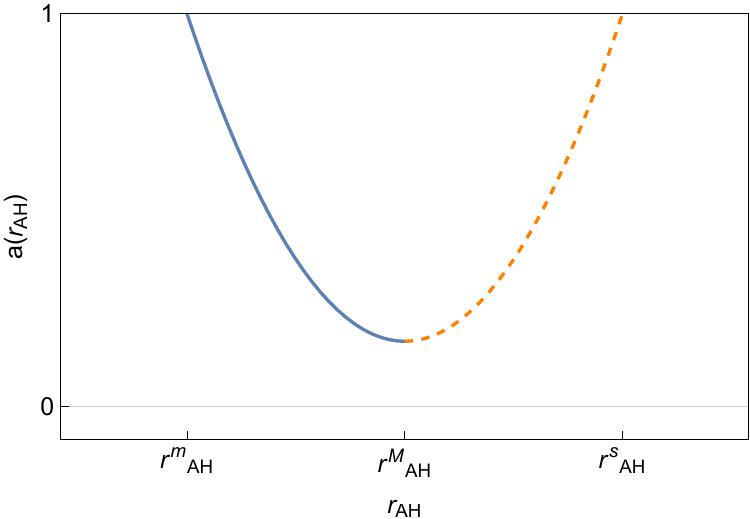}
    \caption{Qualitative behavior of the scale factor as a function of the AH size. Since the behavior is non-monotonic, we can invert this to obtain $\rAH = \rAH(a)$ only in the range $\rAH^m \leq \rAH \leq \rAH^M$ (blue solid curve), while the range $\rAH^M \leq \rAH \leq \rAH^s$ (orange dashed line) is excluded.}
    \label{fig:AHQualitative}
\end{figure}

This indicates that the function $a = a(\rAH)$ defined  by \cref{ApparentHorizonFinalEq1} cannot be monotonic, but it must have at least one minimum at $\rAH= \rAH^M$. Consequently, $a(\rAH)$ cannot be inverted over the whole range $\rAH \in (\rAH^m, \rAH^s)$. Assuming that there are no further extrema, $a(\rAH)$ can be inverted within the region $\rAH^m \leq \rAH\leq \rAH^M$. The qualitative behavior of $a(\rAH)$ is shown in \cref{fig:AHQualitative}.
Notice that $\rAH^M$ depends both on black-hole hairs $M$, $\ell$ and on the cosmological parameters $H_0$, $\Omega$ and $w$. Thus, $\rAH^M$ represents an upper bound on the size of the AH that can be attained by a black hole in a FLRW universe characterized by these cosmological parameters.

Following the reasoning above, \cref{ApparentHorizonFinalEq1} also determines a threshold redshift value $z_M$ beyond which the AH cannot form. If a black hole forms at $z_i<z_M$, the AH appears together with the EH. The size of the AH is maximal at $z_i$ and decreases monotonically until $z=0$, where it attains its minimum value, which is always larger than $\rH$. However, if the black hole forms at $z_i>z_M$, its AH appears only later, once $z$ crosses $z_M$. This peculiar and intriguing feature seems to arise as a consequence of the coupling between a local compact object and the large-scale, cosmological dynamics. As we will see in the next section through explicit models, astrophysical black holes -- both stellar and supermassive -- typically form at $z_i<z_M$. The case where $z_i>z_M$ could, instead, be relevant for primordial black holes.

The behavior of $a(\rAH)$ described above also holds when the function defined in \cref{ApparentHorizonFinalEq1} has more than one extremum. In such cases, $\rAH^M$ corresponds to the first minimum. This behavior remains consistently true even when considering $-1 \leq w \leq-1/3$. While the denominator on the right hand side of \cref{ApparentHorizonFinalEq1} can become negative in this range, this occurs when $\rAH \alpha'$ is relatively small, i.e., at values of $\rAH \gg \rAH^M $.

\section{Cosmological embedding of the Fan $\&$ Wang and Hayward black holes }
\label{sect:4}

In this section, we apply the general method discussed in the previous section to two specific cases of interest: the Fan $\&$ Wang and Hayward black holes. Their metric functions are given by $f_\text{FW}=1-\frac{2GM r^2}{(r + \ell)^3}$ and $f_\text{H} = 1-\frac{2GM r^2}{r^3 + \ell^3}$, respectively, where $M$ is the ADM mass and $\ell$ is an additional parameter -- possibly related to quantum effects -- responsible for the smearing of the singularity (see Ref.~\cite{Cadoni:2022chn}).
The corresponding cosmologically-embedded solutions \eqref{generalmetric} can be easily written using \cref{betafunction,ffunction}.

For the Fan $\&$ Wang black hole we have 
\begin{equation}
\begin{split}
    \dd s^2_\text{FW} &= a^2 \left[-e^{\alpha_\text{FW}} \dd t^2 + e^{-\alpha_\text{FW}} a^{-k_\text{FW}} \dd r^2 + \dd \Omega^2 \right]\, ;\\
    \alpha_\text{FW} (r) & = \ln \left[1-\frac{2GM r^2}{(r + \ell)^3}\right]\, ; \\
    k_\text{FW}(r)  &= \frac{2 G M r^2 (r-2\ell)}{(\ell+r)\left[(r+\ell)^3 -2GM r^2 \right]}\, ,
\end{split}
\label{CEFW}
\end{equation}
while for the Hayward one we get
\begin{equation}
\begin{split}
    \dd s^2_\text{H} &= a^2 \left[-e^{\alpha_\text{H}} \dd t^2 + e^{-\alpha_\text{H}} a^{-k_\text{H}} \dd r^2 + \dd \Omega^2 \right]\, ;\\
    \alpha_\text{H} (r) & = \ln \left[1-\frac{2GM r^2}{r^3 + \ell^3}\right]\, ; \\
    k_\text{H}(r)  &= \frac{2GM r^2 (r^3 - 2\ell^3)}{(r^3 + \ell^3) \left[r^3 + \ell^3 - 2 GMr^2 \right]}\, .
\end{split}
\label{CEH}
\end{equation}
Any local observer, on time scales such that $\dot a\sim 0$, perceives the geometry of a corresponding static black hole. In particular, the local observer will detect an outer EH at $r=\rH$, along with the causal structure discussed in~\cite{Cadoni:2022chn}. However, in comoving coordinates, the relevant causal surface is the AH described by \cref{ApparentHorizonFinalEq}. 
The Misner-Sharp mass $M_\text{MS}$ associated with these solutions can also be calculated. This can be done either from the field equations (see Ref.~\cite{Cadoni:2023lqe}) or by using \cref{MSmassSchwarzschildcoordinates} (see Ref.~\cite{Cadoni:2023lum}). In both approaches, the mass shift formula \eqref{mshift} is recovered, indicating that the mass of the Fan $\&$ Wang and Hayward black holes are cosmologically coupled.

\subsection{Apparent horizon for the  Fan $\&$ Wang and Hayward black holes}

\Cref{ApparentHorizonFinalEq} must be solved numerically. The numerical procedure consists of selecting the black-hole metric to embed -- for instance the Fan $\&$ Wang or Hayward metrics considered here -- and the cosmological-background parameters $w$ and $\Omega$. Then, the redshift-axis is discretized and, for each point along it, \cref{ApparentHorizonFinalEq} is solved numerically to find the corresponding value of $\rAH(z)$. 

We choose to measure $\rAH$ in gravitational radii $GM/c^2$ (for the numerical scheme, we reinstate the units of $c$). On the left hand side of \cref{ApparentHorizonFinalEq}, we have $(H_0^2/c^2) \Omega \rAH^2 e^{-2\alpha}$. Since all length scales are converted to $GM/c^2$, we must specify a test mass for the embedded object. We choose $M = 10^6 \, M_\odot$, corresponding to a typical mass of a supermassive black hole, such as SgrA$^\ast$ at the center of our galaxy. Nonetheless, we expect our results to be qualitatively valid for astrophysical black holes of any mass. 
% Therefore, in the left hand side of \cref{ApparentHorizonFinalEq}, we have
% \begin{equation}
%     \frac{H_0^2}{c^2} \Omega \, \rAH^2 \, e^{-2\alpha} = \frac{H_0^2}{c^2} \, \left(\frac{GM}{c^2} \right)^2 \, \Omega \left(\frac{\rAH}{GM/c^2} \right)^2 \, e^{-2\alpha}\, .
% \end{equation}
%
Adopting also $H_0 = 67.7 \, \text{km} \cdot \text{s}^{-1} \cdot \text{Mpc}^{-1}$, we have
\begin{equation}
    \frac{H_0^2}{c^2} \, \left(\frac{GM}{c^2} \right)^2 \simeq 9.70444 \cdot 10^{-16}\, .
    \label{numericalvalues}
\end{equation}
To mitigate numerical issues caused by the small values in \cref{numericalvalues}, rather than solving \cref{ApparentHorizonFinalEq}, we take the logarithm of both sides
\begin{widetext}
    \begin{equation}
    \ln \left[\frac{H_0^2}{c^2} \, \left(\frac{GM}{c^2} \right)^2 \, \Omega  \right] + 2 \ln \left(\frac{\rAH}{GM/c^2} \right) - 2\alpha = (\rAH \alpha ' + 1 + 3w) \ln\left(\frac{1}{1+z} \right)\, .
    \label{AHfornumericalsolution}
\end{equation}
\end{widetext}
In the following, we consider two cases relevant for cosmologically-embedded  black holes:
\begin{itemize}
    \item Black holes in a dust-dominated cosmological background, for which $\Omega = \Omega_0 \simeq 0.3$~\cite{Planck:2018vyg} and $w = 0$ in \cref{AHfornumericalsolution}. We consider the range of redshift $z \in [0.6, 100]$. 
    \item Black holes in a cosmological-constant-dominated background, for which $\Omega = \Omega_{\Lambda 0} \simeq 0.7$~\cite{Planck:2018vyg} and $w = -1$ in \cref{AHfornumericalsolution}. We focus on the redshift range $z \in [0.005, 0.6]$, which corresponds to the era dominated by the cosmological constant, based on current observations.
\end{itemize}

\subsection*{Apparent horizon of the embedded Fan $\&$ Wang metric}

\begin{figure}
\centering
\subfigure[]{\includegraphics[width=0.45\textwidth]{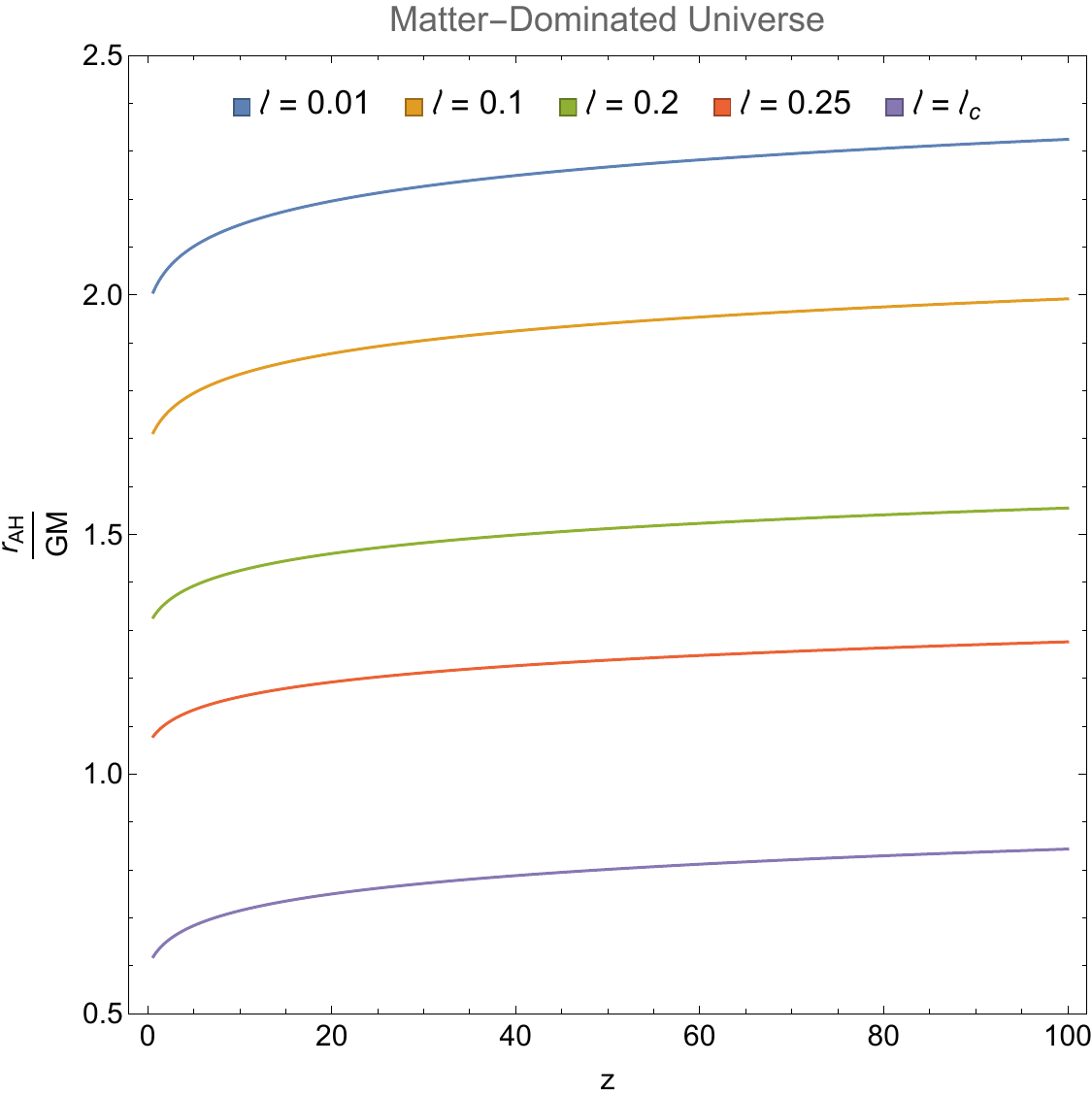}}
\hspace{0.55 cm}
\subfigure[]{\includegraphics[width=0.45\textwidth]{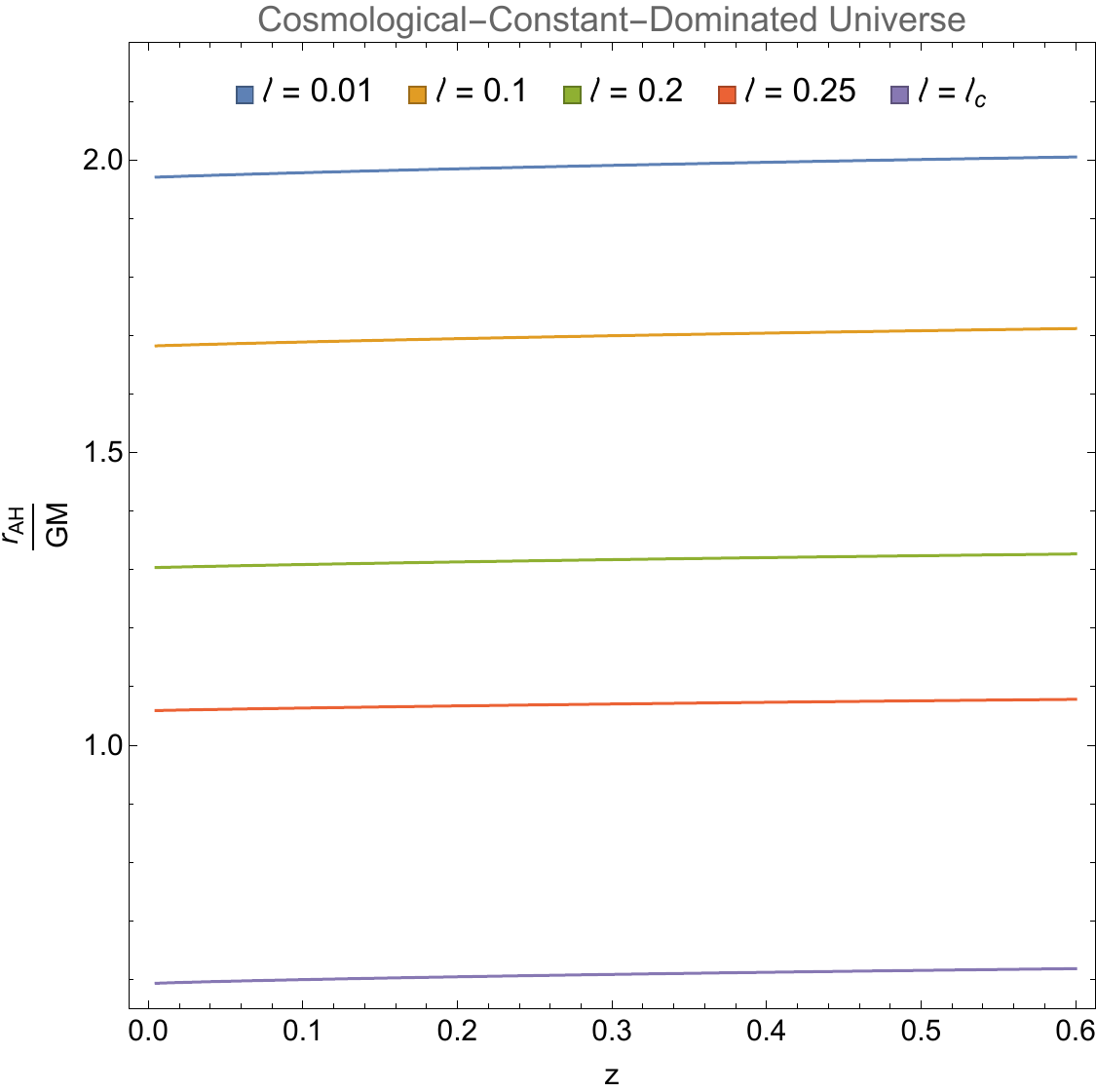}}
\caption{{\bf Figure (a):} apparent horizon of the Fan $\&$ Wang cosmologically embedded black hole as a function of the redshift in a matter-dominated universe. {\bf Figure (b):} apparent horizon of the Fan $\&$ Wang embedded black hole as a function of the redshift in a cosmological-constant-dominated universe.}
\label{fig:FW}
\end{figure}

We begin by considering the exact solution \eqref{CEFW} describing the cosmological embedding of the Fan $\&$ Wang black hole and numerically solve \cref{AHfornumericalsolution} for different values of the hair $\ell$ in both a dust-dominated and a cosmological-constant dominated universe. All chosen values of $\ell$ are consistent with the presence of horizons in the static case, with $\ell_c$ corresponding to the extremal black-hole configuration.
The results are shown in \cref{fig:FW}. 
The behaviour of $\rAH(z)$ fully confirms the general results derived in \cref{sect:31}: $\rAH(z)$ is always a monotonically increasing function of the redsfhift. The minimum values $\rAH^m$, occurs at $z=0$ and is consistently larger than the radius $\rH$ of the EH of a static black hole with the same parameters $M$ and $\ell$.
The maximum value of $\rAH(z)$ is always reached at the time of its formation, i.e., by $\rAH(z_i)$. This is because $z_i$ is always significantly smaller than the threshold value $z_M$ for AH formation (see \cref{sect:31}). Indeed, we have checked that $z_M \sim 10^{11}$ for both matter- and cosmological-constant-dominated universes, meaning that the condition $z_i \ll z_M$ is always satisfied. For astrophysical objects forming at redshifts $z \sim 10-20$, the AH is always of the order of $GM$.

\subsection*{Apparent horizon of the embedded Hayward metric}

\begin{figure}
\centering
\subfigure[]{\includegraphics[width=0.45\textwidth]{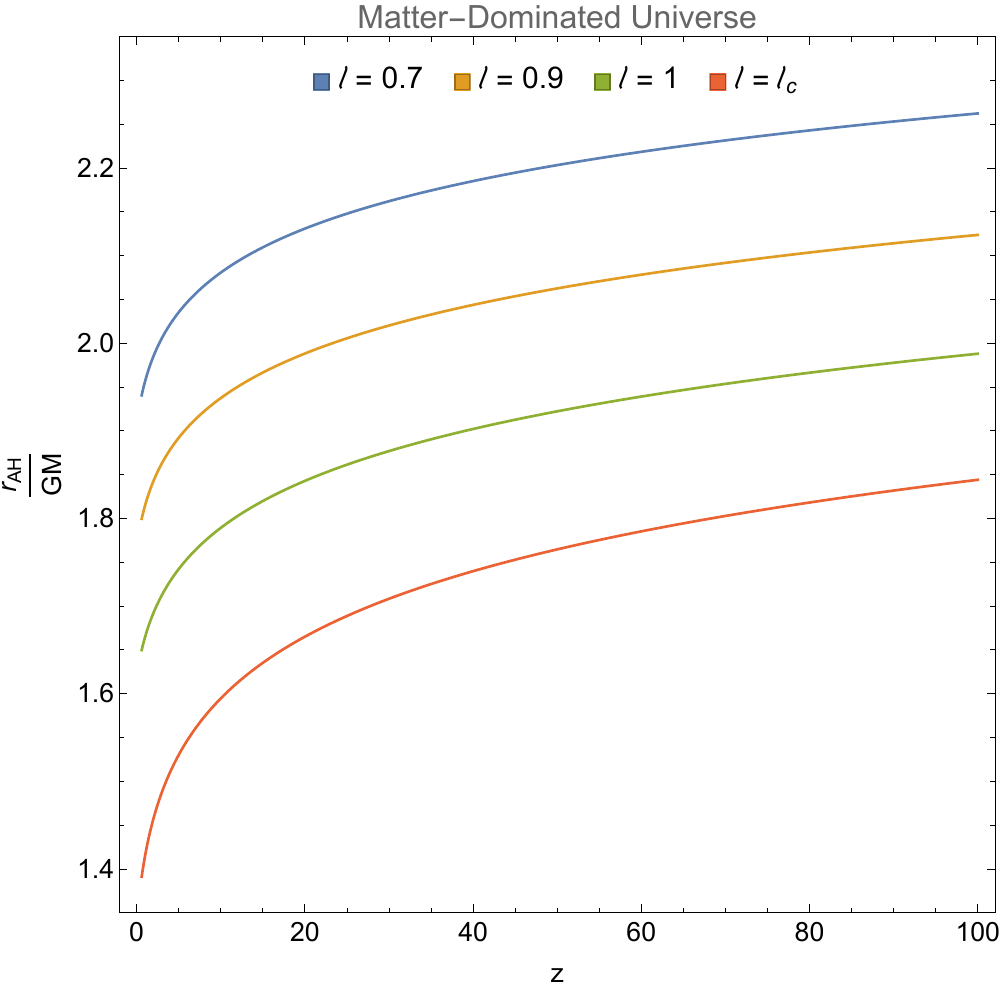}}
\hspace{0.55 cm}
\subfigure[]{\includegraphics[width=0.45\textwidth]{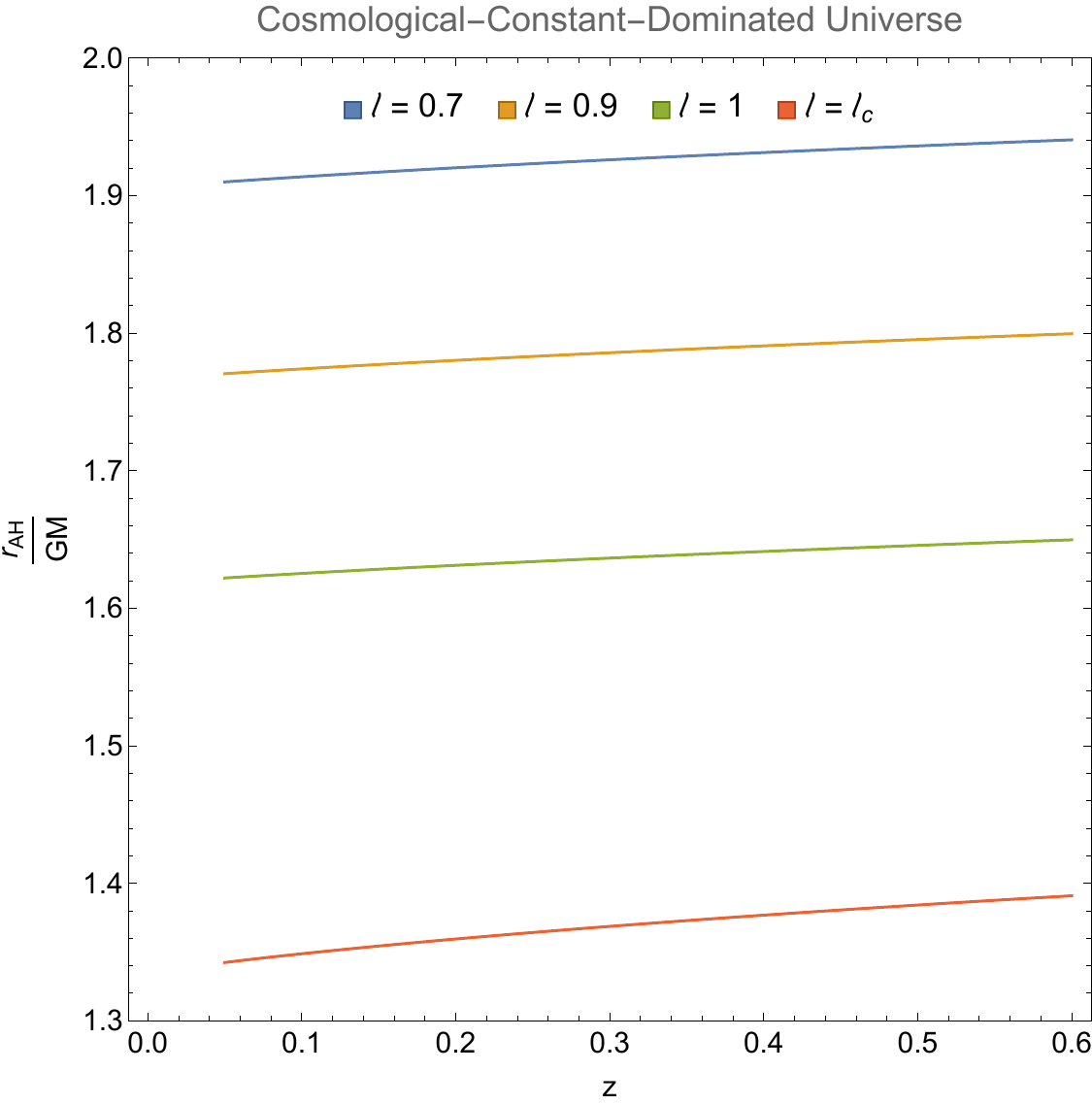}}
\caption{{\bf Figure (a):} apparent horizon of the Hayward embedded black hole as a function of the redshift in a matter-dominated universe. {\bf Figure (b):} apparent horizon of the Hayward embedded black hole as a function of the redshift in a cosmological-constant-dominated universe.}
\label{fig:H}
\end{figure}

Next, we examine the cosmological embedding of the Hayward black hole, as described by \cref{CEH}. Again, we numerically solve \cref{AHfornumericalsolution}  for different values of the hair $\ell \leq \ell_c$, again considering both matter-dominated and cosmological-constant dominated universe. The results are shown in \cref{fig:H}.

The time-evolution of the apparent horizon for the cosmologically embedded Hayward black hole fully confirms the results from \cref{sect:31}. As in the Fan $\&$ Wang case, $\rAH(z)$ is a monotonically increasing function of the redshift. The minimum values of $\rAH^m$ are consistently larger than those of the EHs of the corresponding static black holes. Similarly to the case of the previous section, the inequality $z_i \ll z_M$ holds for both matter-dominated and cosmological-constant dominated universes (also here, $z_M \sim 10^{11}$). Again, for astrophysical black holes forming at $z \sim 10-20$, the AH is always of the order of $GM$.

\section{Cosmologically coupled  black holes: role of the event and apparent horizons}
\label{sect:5}
We are now in a position to answer the questions raised in the introduction. The first question concerns the relationship between the existence of an AH and the cosmological mass shift. In the previous sections, we have demonstrated that our broad class of non-singular black holes, including the Fan $\&$ Wang and Hayward black holes as particular cases, always possesses a well-defined AH. Furthermore, it has been previously established~\cite{Cadoni:2023lqe, Cadoni:2023lum} that the mass of non-singular black holes is cosmologically coupled, following the mass shift formula given by \cref{mshift}. 

However, it is also known that cosmologically embedded solutions exist, such as the McVittie embedding of the Schwarzschild black hole, which possess an apparent horizon, but do not exhibit cosmological mass coupling (i.e., they do not follow \cref{mshift}, but rather \cref{msf} with $k=0$)~\cite{Cadoni:2023lqe}.
This implies that the presence of an AH is not necessarily indicative of cosmological mass coupling. 

In other words, there are two versions of cosmological coupling for black holes: a “weak” version, which implies only the existence of an AH, and a “strong” version, which involves the cosmological coupling of the black hole’s mass. At present, it remains unclear whether cosmological embedding of local gravitational structures can occur without both an AH and cosmological mass coupling. 

The distinction between the weak and strong versions of the cosmological coupling aligns with the results of Ref.~\cite{Faraoni:2024ghi}. A black hole embedded in a cosmological, i.e., time-dependent, background cannot possess a global EH, only an AH. However, the existence of an AH does not necessarily imply a cosmological mass shift, as the latter requires additional conditions beyond mere cosmological embedding.

The second question concerns the relationship between the event and apparent horizons. Unlike in the case of a collapsing body, where the formation of an AH signals the eventual formation of an EH, we find no evidence that their existence is correlated in our scenario. In fact, the AH radius is always larger than that of the EH. Moreover, when a black hole forms at a redshift $z_i > z_M$, the AH is absent, even though the local black-hole solution still has an EH.

These features arise because, in a fully cosmological regime, the concepts of event and apparent horizons are not compatible, as also suggested by the findings of Ref.~\cite{Faraoni:2024ghi}. In a cosmological context, the EH is meaningful only for a local observer using local coordinates and over timescales where the black-hole solution can be considered fully static.

On the other hand, the time evolution of the AH in our CEBHSs seems to be a typical feature of cosmological embedding, reflecting the coupling between the local black-hole and the large-scale cosmological dynamics. For example, the monotonic behavior of the AH radius $\rAH$ with the redshift is naturally expected for a black hole coupled to an expanding universe. It would be also interesting to explore whether AHs might arise in cases where local structures without an EH, such as galaxies, are cosmologically embedded (see Ref.~\cite{Cadoni:2024jxy}). 

\medskip

\section*{Acknowledgments}
We are grateful to Valerio Faraoni and Massimiliano Rinaldi for useful discussions. 

\bibliography{refs}
\end{document}